\documentclass[sigconf]{acmart}

\usepackage{booktabs} 

\usepackage{CJKutf8}

\setcopyright{rightsretained}
\usepackage{comment}
\setcopyright{rightsretained}
\usepackage{booktabs} 

\usepackage{helvet}  
\usepackage{courier}  
\usepackage{url}  
\usepackage{graphicx}  
\usepackage{subfigure}
\usepackage{courier}
\usepackage{color, colortbl}
\usepackage{verbatim}
\usepackage[linesnumbered,ruled,vlined]{algorithm2e}
\usepackage{amssymb,amsmath}
\usepackage{booktabs}
\usepackage{mathrsfs}
\usepackage{makecell}

\DeclareMathOperator*{\argmin}{argmin} 
\DeclareMathOperator*{\argmax}{argmax}

\acmDOI{10.475/123_4}

\acmISBN{123-4567-24-567/08/06}

\acmConference[WOODSTOCK'97]{ACM Woodstock conference}{July 1997}{USA} 
\acmYear{1997}
\copyrightyear{2018}

\acmArticle{4}
\acmPrice{15.00}

\begin{document}
\title{Leveraging Long and Short-term Information in Content-aware Movie
Recommendation}


\author{Wei Zhao}
\affiliation{%
  \institution{SIAT, Chinese Academy of Sciences}
}
\email{wei.zhao@siat.ac.cn}
  
\author{Haixia Chai, Benyou Wang}
\affiliation{%
  \institution{Tencent}
}
\email{{haixiachai,wabywang}@tencent.com}

\author{Jianbo Ye}
\affiliation{%
  \institution{Pennsylvania State University}
}
\email{jxy198@ist.psu.edu}

\author{Min Yang}
\authornote{Min Yang is the corresponding author.}
\affiliation{%
  \institution{SIAT, Chinese Academy of Sciences}
}
\email{min.yang@siat.ac.cn}

\author{Zhou Zhao}
\affiliation{%
  \institution{Zhejiang University}
}
\email{zhaozhou@zju.edu.cn}

\author{Xiaojun Chen}
\affiliation{%
  \institution{Shenzhen University}
}
\email{xjchen@szu.edu.cn}


\begin{abstract}
Movie recommendation systems provide users with ranked lists of movies based on individual's preferences and constraints. 
Two types of models are commonly used to generate ranking results: long-term models and session-based models. While long-term models represent the interactions between users and movies that are supposed to change slowly across time, session-based models encode the information of users' interests and changing dynamics of movies' attributes in short terms. 
In this paper, we propose an \emph{LSIC} model, leveraging \textbf{L}ong and \textbf{S}hort-term \textbf{I}nformation in \textbf{C}ontent-aware movie recommendation using adversarial training. 
In the adversarial process, we train a generator as an agent of reinforcement learning which recommends the next movie to a user sequentially. We also train a discriminator which attempts to distinguish the generated list of movies from the real records. The poster information of movies is integrated to further improve the performance of movie recommendation, which is specifically essential when few ratings are available. The experiments demonstrate that the proposed model has robust superiority over competitors and sets the state-of-the-art. We will release the source code of this work after publication. 


\end{abstract}
%
%
\ccsdesc[100]{Recommender system~Movie Recommendation}


\keywords{Movie recommendation, Adversarial learning, Content-aware recommendation}

\maketitle
\renewcommand{\shortauthors}{Zhao et al.}

\section{Introduction}
With the sheer volume of online information, much attention has been given to data-driven recommender systems.  Those systems automatically guide users to discover products or services respecting their personal interests from a large pool of possible options. 
Numerous recommendation techniques have been developed. Three main categories of them are: collaborative filtering methods, content-based methods and hybrid methods~ \cite{bobadilla2013recommender,lu2015recommender}. In this paper, we aim to develop a method producing a ranked list of $n$ movies to a user at a given moment (top-$n$ movie recommendation) by exploiting both historical user-movie interactions and the content information of movies.

Matrix factorization (MF) \cite{koren2009matrix} is one of the most successful techniques in the practice of recommendation due to its simplicity, attractive accuracy and scalability. It has been used in a broad range of applications such as recommending movies, books, web pages, relevant research and services. The matrix factorization technique is usually effective because it discovers the latent features underpinning the multiplicative interactions between users and movies. Specifically, it models the user preference matrix approximately as a product of two lower-rank latent feature matrices representing user profiles and movie profiles respectively. 

Despite the appeal of matrix factorization, this technique does not explicitly consider the temporal variability of data~\cite{wu2017recurrent}. Firstly, the popularity of an movie may change over time. For example, movie popularity booms or fades, which can be triggered by external events such as the appearance of an actor in a new movie. Secondly, users may change their interests and baseline ratings over time. For instance, a user who tended to rate an average movie as ``4 stars'', may now rate such a movie as ``3 stars''. Recently, recurrent neural network (RNN)~\cite{hochreiter1997long} has gained significant attention by considering such temporal dynamics for both users and movies and achieved high recommendation quality~\cite{wu2016joint,wu2017recurrent}. The basic idea of these RNN-based methods is to formulate the recommendation as a sequence prediction problem. They take the latest observations as input, update the internal states and make predictions based on the newly updated states. As shown in~\cite{devooght2016collaborative}, such prediction based on short-term dependencies is likely to improve the recommendation diversity.

More recent work~\cite{wu2017recurrent} reveals that matrix factorization based and RNN-based recommendation approaches have good performances for the reasons that are complementary to each other. Specifically, the matrix factorization recommendation approaches make movie predictions based on users' long-term interests which change very slowly with respect to time. On the contrary, the RNN recommendation approaches predict which movie will the user consume next, respecting the dynamics of users' behaviors and movies' attributes in the short term. It therefore motivates us to devise a joint approach that takes advantage of both matrix factorization and RNN, exploiting both long-term and short-term associations among users and movies.


Furthermore, most existing recommender systems  take into account only the users' past behaviors when making recommendation.  compared with tens of thousands of movies in the corpus, the historical rating set is too sparse to learn a well-performed model.  It is desirable to exploit the content information of movies for recommendation.  For example,  movie
posters reveal a great amount of information
to understand movies and users, as demonstrated in~\cite{zhao2016matrix}. 
Such a poster is usually the first contact that a user has with a movie, and plays 
an essential role in the  user's decision to watch it or not. 
When a user is watching the movie presented in cold, blue and
mysterious visual effects, he/she may be interested in receiving
recommendations for movies with similar styles, rather
than others that are with the same actors or subject~\cite{zhao2016matrix}. 
These visual features of  movies are usually captured by the corresponding posters.




In this paper, we propose a novel LSIC model, which leverages \textbf{L}ong and \textbf{S}hort-term \textbf{I}nformation in \textbf{C}ontent-aware movie recommendation using adversarial training. 
The LSTC model employs an adversarial framework to combine the MF and RNN based models for the top-$n$ movie recommendation, taking the best of each to improve the final recommendation performance. In the adversarial process, we simultaneously train two models: a generative model $G$ and a discriminative model $D$. In particular, the generator $G$ takes the user $u_{i}$ and time $t$ as input, and predicts the recommendation list for user $i$ at time $t$ based on the historical user-movie interactions. We implement the discriminator $D$ via a siamese network that incorporates long-term and session-based ranking model in a pair-wise scenario. The two point-wise networks of siamese network share the same set of parameters. The generator $G$ and the discriminator $D$ are optimized with a minimax two-player game. The discriminator $D$ tries to distinguish the real high-rated movies in the training data from the  recommendation list generated by the generator $G$, while the training procedure of generator $G$ is to maximize the probability of $D$ making a mistake. Thus, this adversarial process can eventually adjust $G$ to generate plausible and high-quality recommendation list. 
In addition, we integrate poster information of movies to further improve the performance of movie recommendation, which is specifically essential when few ratings are available. 

We summarize our main contributions as follows:
\begin{itemize}
\item To the best of our knowledge, we are the first to use GAN framework to leverage the MF and RNN approaches for top-$n$ recommendation. This joint model adaptively adjusts how the contributions of the long-term and short-term information of users and movies are mixed together. 
\item We propose hard and soft mixture mechanisms to integrate MF and RNN. We use the hard mechanism to calculate the mixing score straightforwardly and explore several soft mechanisms to learn the temporal dynamics with the help of the long-term profiles.
\item Our model uses reinforcement learning to optimize the generator $G$ for generating highly rewarded recommendation list. Thus, it effectively bypasses the non-differentiable task metric issue by directly performing policy gradient update. 
\item We automatically crawl the posters of the given movies, and explore the potential of integrating poster information to improve the accuracy of movie recommendation. The release of the collected posters would push forward the research of integrating content information in movie recommender systems. 
\item To verify the effectiveness of our model, we conduct extensive experiments on two widely used real-life datasets: Netflix Prize Contest data and Movielens data. The experimental results demonstrate that our model consistently outperforms the state-of-the-art methods. 
\end{itemize}

The rest of the paper is organized as follows. In Section 2, we review the related work on recommender systems. Section 3 presents the proposed adversarial learning framework for movie recommendation in details. In Section 4, we describe the experimental data, implementation details, evaluation metrics and baseline methods. The experimental results and analysis are provided in Section 5. Section 6 concludes this paper.

\section{Related Work}
Recommender system is an active research field.
The authors of~\cite{bobadilla2013recommender,lu2015recommender}
describe most of the existing techniques for recommender systems.
In this section, we briefly review the following major approaches
for recommender systems that are mostly related to our work. 
\paragraph{Matrix factorization for recommendation}
Modeling the long-term interests of users, the matrix factorization method and
its variants have grown to become dominant in the literature \cite{rennie2005fast,koren2008factorization,koren2009matrix,hernando2016non,he2016fast}.
In the standard matrix factorization, the recommendation task can
be formulated as inferring missing values of a partially observed
user-item matrix \cite{koren2009matrix}.
The Matrix Factorization techniques are effective because they are designed to discover the latent
features underlying the interactions between users and items. 
\citeauthor{srebro2005maximum} \shortcite{srebro2005maximum} suggested the Maximum Margin Matrix
Factorization (MMMF), which used low-norm instead of low-rank factorizations.
\citeauthor{mnih2008probabilistic} \shortcite{mnih2008probabilistic} presented the Probabilistic
Matrix Factorization (PMF) model that characterized the user preference
matrix as a product of two lower-rank user and item matrices. The
PMF model was especially effective at making better predictions for
users with few ratings. \citeauthor{he2016fast} \shortcite{he2016fast} proposed a new
 MF method which considers the implicit feedback for on-line recommendation.
In \cite{he2016fast}, the weights of the missing data were assigned based on the popularity of items. 
To exploit the content of items and solve the sparse issue in recommender systems, \cite{zhao2016matrix} presented the model for movie recommendation using additional visual features (e.g.. posters and still frames) to better understand movies, and further improved the performance of movie recommendation.

\paragraph{Recurrent neural network for recommendation}

These traditional MF methods for recommendation systems are based on
the assumption that the user interests and movie attributes are near static,
which is however not consistent with reality. \citeauthor{koren2010collaborative} \shortcite{koren2010collaborative}
discussed the effect of temporal dynamics in recommender systems
and proposed a temporal extension of the SVD++ (called TimeSVD++)
to explicitly model the temporal bias in data. However, the features
used in TimeSVD++ were hand-crafted and computationally expensive to obtain. Recently,
there have been increasing interests in employing recurrent neural
network to model temporal dynamic in recommendation systems. For example, \citeauthor{hidasi2015session} \shortcite{hidasi2015session} applied recurrent
neural network (i.e. GRU) to session-based recommender systems. This work
treats the first item a user clicked as the initial input of GRU.
Each follow-up click of the user would then trigger a recommendation
depending on all of the previous clicks. \citeauthor{wu2016recurrent} \shortcite{wu2016recurrent}
proposed a recurrent neural network to perform the time heterogeneous
feedback recommendation. \citeauthor{wu2017recurrent} \shortcite{wu2017recurrent} used LSTM autoregressive model for the user and movie dynamics and employed matrix factorization to model
the stationary components that encode fixed properties. Different from their work, 
we use GAN framework to leverage the MF and RNN approaches for top-$n$ recommendation, 
aiming to generate plausible and high-quality recommendation lists.  
To address the cold start problem in recommendation,  \cite{cui2016visual}  presented a visual and textural recurrent neural network (VT-RNN), which simultaneously learned the sequential latent vectors of user’s interest and captured the content-based representations that contributed
to address the cold start.
\paragraph{Generative adversarial network for recommendation}

In parallel, previous work has demonstrated the effectiveness of generative
adversarial network (GAN) \cite{goodfellow2014generative} in various
tasks such as image generation \cite{reed2016generative,arjovsky2017wasserstein},
image captioning \cite{chen2017show}, and sequence generation\cite{yu2017seqgan}.
The most related work to ours is \cite{wang2017irgan}, which proposed
a novel IRGAN mechanism to iteratively optimize a generative retrieval
component and a discriminative retrieval component. IRGAN
reported impressive results on the tasks of web search, item recommendation,
and question answering. Our approach differs from theirs
in several aspects. First, we combine the MF approach and the RNN
approach with GAN, exploiting the performance contributions of both approaches. 
Second, IRGAN does not attempt to estimate the future behavior since the experimental data is split randomly in their setting. In fact, they use future trajectories to infer the historical records, which seems not useful in real-life applications. 
Third, we incorporate poster information of movies to deal with the cold-start issue and boost the recommendation performance. 

\section{Our Model}

\begin{table}[t]
\small
	\centering
	\caption{Notation list. We use superscript $u$ to annotate parameters related to a user, and superscript $m$ to annotate parameters related to a movie.}
	\label{tab:symbol}
	\begin{tabular}{l|l}
		\toprule
		\midrule

         $R$ & the user-movie rating matrix\\
        $U$, $M$ &  the number of users and movies\\
        $r_{ij}$ & rating score of user $i$ on movie $j$\\
        $r_{ij,t}$ & rating score of user $i$ on movie $j$ at time $t$\\
        $\mathbf e^{u}_{i}$ & MF user factors for user $i$\\
        $\mathbf e^{m}_{j}$ & MF movie factors for movie $j$\\
        $b^{u}_{i}$ & bias of user $i$ in MF and RNN hybrid calculation \\
        $b^{m}_{j}$ & bias of movie $j$ in MF and RNN hybrid calculation\\
        $\mathbf h^{u}_{i,t}$ & LSTM hidden-vector at time $t$ for user $i$\\
        $\mathbf h^{m}_{j,t}$ & LSTM hidden-vector at time $t$ for movie $j$\\
        $\mathbf z^{u}_{i,t}$ & the rating vector of user $i$ at time $t$ (LSTM input)\\
        $\mathbf z^{m}_{j,t}$ & the rating vector of movie $j$ at time $t$ (LSTM input)\\
        $\alpha^{i}_{t}$ & attention weight of user $i$ at time $t$\\
        $\beta^{j}_{t}$ & attention weight of movie $j$ at time $t$\\
        $m_{+}$ & index of a positive (high-rating) movie drawn from the  \\
        & entire positive  movie set  $\mathcal M_+$ \\
        $m_{-}$ & index of a negative (low-rating) movie randomly chosen from the \\
        &entire negative  movie set $\mathcal M_-$\\
        $m_{g,t}$ & index of an item chosen by generator $G$ at time $t$\\
	\bottomrule
	\end{tabular}
	\vspace{-0.3cm}    
\end{table}

Suppose there is a sparse user-movie rating matrix $R$ that consists of $U$ users and $M$ movies. Each entry $r_{ij,t}$ denotes the rating of user $i$ on movie $j$ at time step $t$. The rating is represented by numerical values from 1 to 5, where the higher value indicates the stronger preference. 
Instead of predicting the rating of a specific user-movie pair as is done in~\cite{adomavicius2005toward,mcnee2006being}, the proposed LSIC model aims to provide users with ranked lists of movies (top-$n$ recommendation) \cite{liu2008eigenrank}. 

\begin{figure*}[h]
\centering
\includegraphics[ width=5.25in]{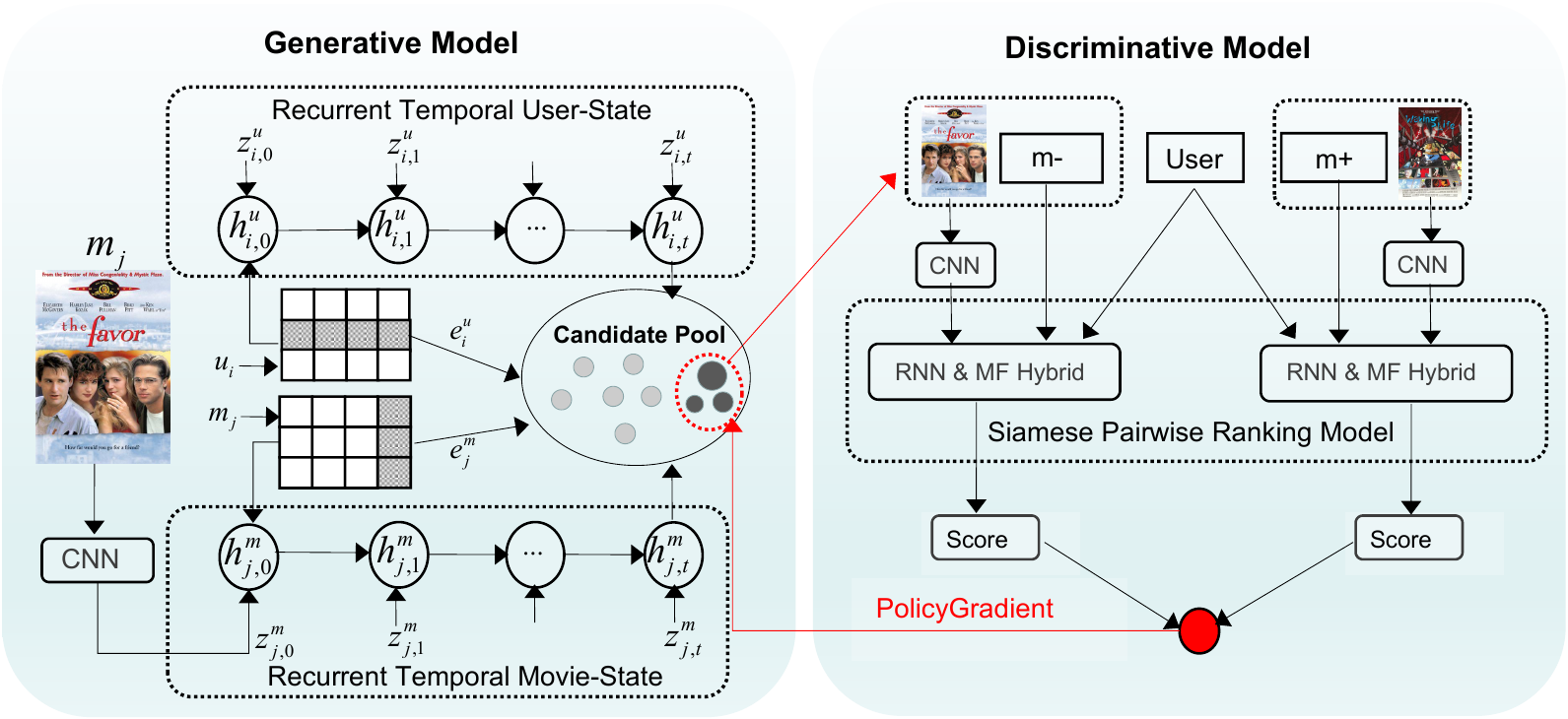}
\caption{The Architecture of Long-term and Session-based Ranking Model with Adversarial Network.
 }
\label{fig:1}
\end{figure*}

In this section, we elaborate each component of LSIC model for content-aware movie recommendation.  
The main notations of this work are summarized in Table \ref{tab:symbol} for clarity. 
The LSTC model employs an adversarial framework to combine the MF and RNN based models for the top-$n$ movie recommendation. The overview of our proposed architecture and its data-flow are illustrated in Figure \ref{fig:1}.  In the adversarial process, we simultaneously train two models: a generative model $G$ and a discriminative model $D$.

\subsection{Matrix Factorization (MF)} 
The MF framework \cite{mnih2008probabilistic} models the long-term states (global information) for both users  ($\mathbf e^u$) and movies ($\mathbf e^m$). In its standard setting, the recommendation task can be formulated as inferring missing values of a partially observed user-movie rating matrix $R$. 
The formulation of MF is given by:
\begin{equation}
\label{eq:mf}
\argmin_{\mathbf e^u,\mathbf{e}^m} \sum_{i,j} I_{ij}\left(r_{ij}-\rho\left((\mathbf e^{u}_i)^T \mathbf e^{m}_{j}\right)\right)^2
+{\lambda^{u}}\|\mathbf{e}^u\|_{F}^2 +{\lambda^{m}}\|\mathbf{e}^m\|_{F}^2
\end{equation}
where $\mathbf{e}^{u}_{i}$ and $\mathbf{e}^{m}_{j}$ represent the user and movie latent factors in the shared $d$-dimension space respectively. $r_{ij}$ denotes the user $i$'s rating on movie $j$. $I_{ij}$ is an indicator function and equals 1 if $r_{ij}>0$, and 0 otherwise. $\lambda^{u}$ and $\lambda^{m}$ are regularization coefficients. The $\rho(\cdot)$ is a logistic scoring function that bounds the range of outputs.

In most recommender systems, matrix factorization techniques \cite{koren2009matrix} recommend movies based on estimated ratings. Even though the predicted ratings can be used to rank the movies, it is known that it does not provide the best prediction for the top-$n$ recommendation. Because minimizing the objective function -- the squared errors -- does not perfectly align with the goal to optimize the ranking order. In this paper, we apply MF for ranking prediction (top-$n$ recommendation) directly, similar to \cite{wang2017irgan}.  
\subsection{Recurrent Neural Network (RNN)} 
The RNN based recommender system focuses on modeling session-based trajectories instead of global (long-term) information \cite{wu2017recurrent}. It predicts future behaviors and provides users with a ranking list given the users' past history. The main purpose of using RNN is to capture time-varying state for both users and movies. 
Particularly, we use LSTM cell as the basic RNN unit. Each LSTM unit at time $t$ consists of a memory cell $c_t$, an input gate $i_t$, a forget gate $f_t$, and an output gate $o_t$. These gates are computed from previous hidden state $\mathbf h_{t-1}$ and the current input $\mathbf x_t$:
\begin{equation}
\begin{aligned}
\left[f_{t},i_{t},o_{t}\right] &= \text{sigmoid}(W[\mathbf h_{t-1}, \mathbf x_{t}])\\
\end{aligned}
\end{equation}

The memory cell $c_t$ is updated by partially forgetting the existing memory and adding a new memory content $\mathbf l_{t}$:
\begin{align}
\mathbf l_{t}&=\tanh(V[\mathbf h_{t-1}, \mathbf x_{t}])\\
\mathbf c_{t}&=f_{t}\odot \mathbf c_{t-1}+i_{t}\odot  \mathbf l_{t}
\end{align}

Once the memory content of the LSTM unit is updated, the hidden state at time step $t$ is given  by:
\begin{equation}
\begin{aligned}
\mathbf h_{t}&=o_{t}\odot \tanh(\mathbf c_{t})
\end{aligned}
\end{equation}
For simplicity of notation, the update of the hidden states of LSTM at time step $t$ is denoted as 
$\mathbf h_{t}={\rm LSTM}(\mathbf h_{t-1}, \mathbf x_{t})$.

Here, we use $\mathbf z^{u}_{i,t} \in \mathbb{R}^{U}$ and $\mathbf z^{m}_{j,t} \in \mathbb{R}^{M}$ to represent the rating vector of user $i$ and movie $j$ given time $t$ respectively. Both $\mathbf z^{u}_{i,t}$ and $\mathbf z^{m}_{j,t}$ serve as the input to the LSTM layer at time $t$ to infer the new states of the user and the movie:
\label{eq:lstm-user}
\begin{align}
\mathbf h^{u}_{i,t} &= {\rm LSTM}(\mathbf h^{u}_{i,t-1},  \mathbf z^{u}_{i,t})\label{eq:lstm-user}\\
\mathbf h^{m}_{j,t} &= {\rm LSTM}(\mathbf h^{m}_{j,t-1}, \mathbf z^{m}_{j,t})\label{eq:lstm-movie}
\end{align}
Here,  $\mathbf h^{u}_{i,t}$ and $\mathbf h^{m}_{j,t}$ denote hidden states for user $i$ and movie $j$ at time step $t$ respectively. 

In this work, we explore the potential of integrating posters of movies to boost the performance of movie recommendation. Inspired by the recent advances of CNNs in computer vision,  the poster is mapped to the same space of the movie  by using a CNN. More concretely, we encode each image into a FC-2k feature vector with Resnet-101 (101 layers), resulting in a 2048-dimensional vector representation. The poster $P_j$ of movie $j$ is only inputted once, at $t=0$, to inform the movie LSTM about the poster content: 

\begin{equation}
\label{eq:lstm-cnn}
\begin{aligned}
\mathbf z^{m}_{j,0}=CNN(P_j). 
\end{aligned}
\end{equation}

\subsection{RNN and MF Hybrid}
The session-based model deals with temporal dynamics of the user and movie states, we further incorporate the long-term preference of users and the fixed properties of movies. To exploit their advantages together, similar to \cite{wu2017recurrent}, we define the rating prediction function as:
\begin{equation}
\begin{aligned}
r_{ij,t} &= g(\mathbf e^{u}_{i}, \mathbf e^{m}_{j}, \mathbf h^{u}_{i,t}, \mathbf h^{m}_{j,t})
\end{aligned}
\end{equation}
where $g(.)$ is a score function, $\mathbf e^{u}_{i}$ and $\mathbf e^{m}_{j}$ denote the global latent factors of user $i$ and movie $j$ learned by Eq.~ (\ref{eq:mf}); $\mathbf h^{u}_{i,t}$ and $\mathbf h^{m}_{j,t}$ denote the hidden states at time step $t$ of two RNNs learned by Eq.~(\ref{eq:lstm-user}) and Eq.~(\ref{eq:lstm-movie}) respectively.
In this work, we study four strategies to calculate the score function $g$, integrating MF and RNN. The details are described below. 



\begin{figure*}
\begin{minipage}{0.25\linewidth}  
	\centerline{\includegraphics[width=4.0cm]{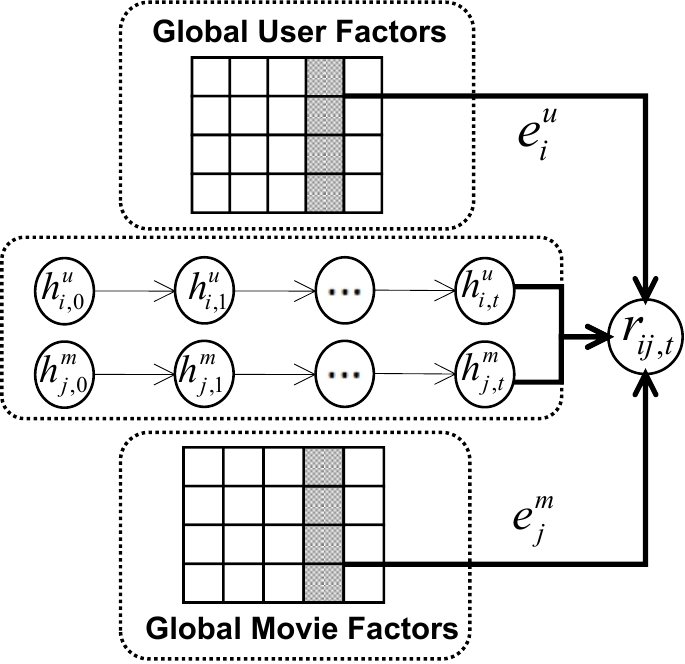}}  
	\centerline{(a) LSIC-V1: Hard mechanism}  
\end{minipage} 
\begin{minipage}{0.24\linewidth}  
	\centerline{\includegraphics[width=4.0cm]{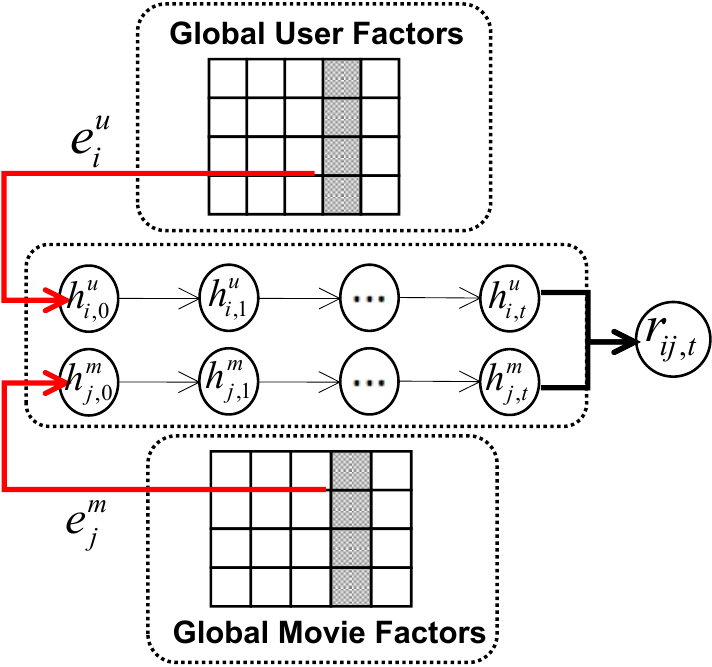}}  
	\centerline{(b) LSIC-V2: Prior initialization}  
\end{minipage}  
\begin{minipage}{0.24\linewidth}  
	\centerline{\includegraphics[width=4.0cm]{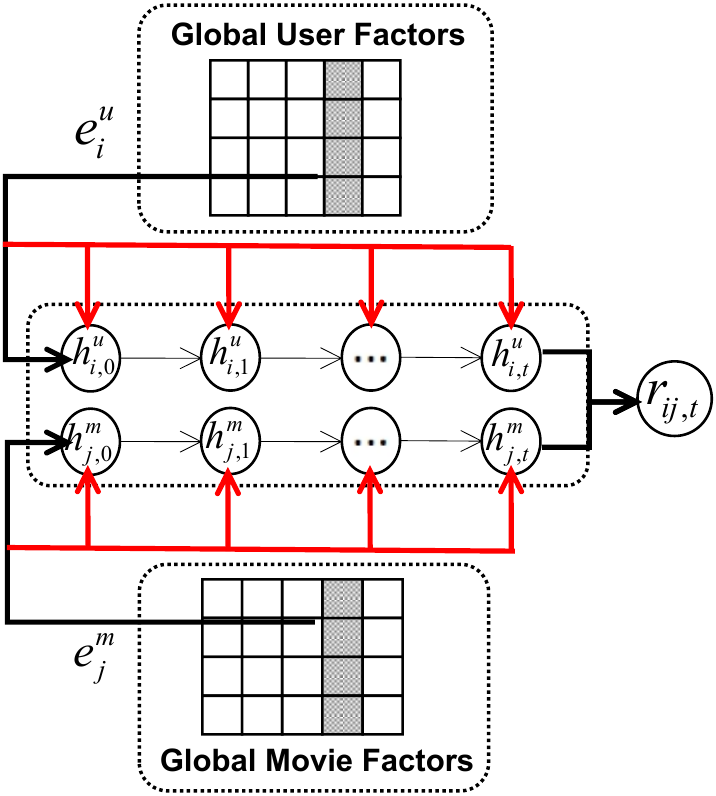}}  
	\centerline{(c) LSIC-V3: Static context}  
\end{minipage}   
\begin{minipage}{0.24\linewidth}  
	\centerline{\includegraphics[width=4.0cm]{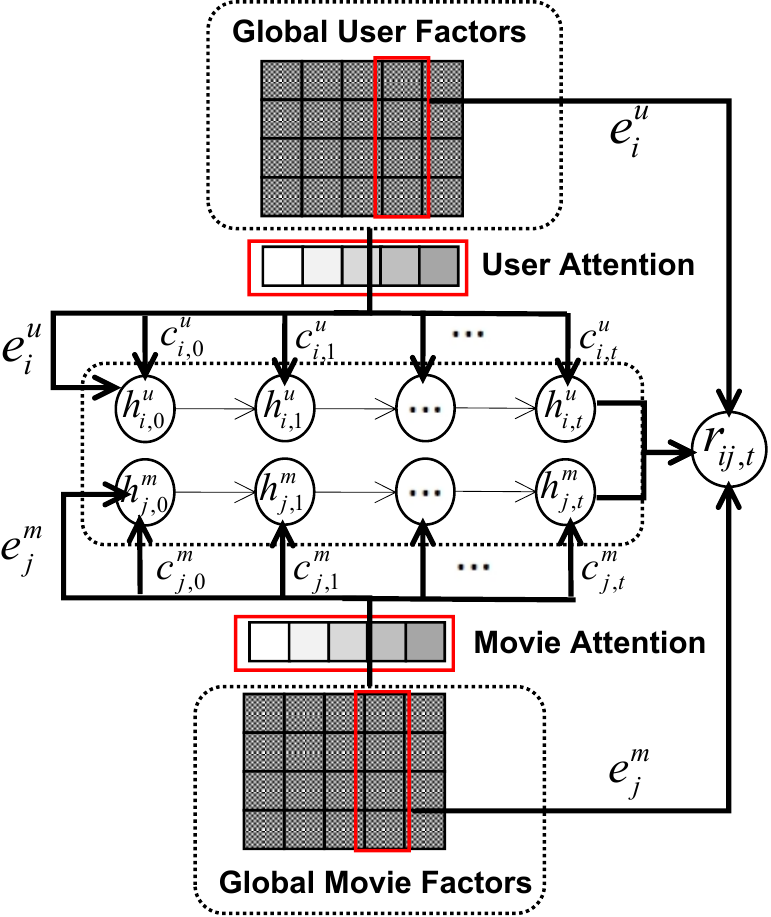}}  
	\centerline{(d) LSIC-V4: Attention model}  
\end{minipage} 
\caption{Four strategies to calculate the score function $g$, integrating MF and RNN.}
 \label{fig:mixture}  
\end{figure*}

\paragraph{LSIC-V1}
This is a hard mechanism, using a simple way to calculate the mixing score from MF and RNN with the following formulation:
\begin{align}
r_{ij,t} &=g(\mathbf e^{u}_{i}, \mathbf e^{m}_{j}, \mathbf h^{u}_{i,t}, \mathbf h^{m}_{j,t})=\frac{1}{1+{\rm exp}(-\mathbf s)} \label{eq:hard}\\
\mathbf s&=\mathbf e^{u}_{i} \cdot \mathbf e^{m}_{j} + \mathbf h^{u}_{i,t} \cdot \mathbf h^{m}_{j,t} +b^{u}_{i}+b^{m}_{j} \label{eq:hard-score}
\end{align}
where $b^{u}_{i}$ and $b^{m}_{j}$ are the biases of user $i$ and movie $j$; $\mathbf h^{u}_{i,t}$ and $\mathbf h^{m}_{j,t}$ are computed by Eq.~(\ref{eq:lstm-user}) and Eq.~(\ref{eq:lstm-movie}). 

In fact, LSIC-V1 does not exploit the global factors in learning the temporal dynamics. In this paper, we also design a soft mixture mechanism and provide three strategies to account for the global factors $\mathbf e^{u}_i$ and $\mathbf e^{m}_j$ in learning $\mathbf h^{u}_{i,t}$ and $\mathbf h^{m}_{j,t}$, as described below (i.e., LSIC-V2, LSIC-V3 and LSIC-V4). 
\paragraph{LSIC-V2}
We use  the latent factors of user $i$ ($\mathbf e^{u}_i$) and movie $j$  ($\mathbf e^{m}_j$) pre-trained by MF model  to initialize the hidden states of the LSTM cells $\mathbf h^{u}_{i,0}$ and $\mathbf h^{m}_{j,0}$ respectively, as depicted in Figure \ref{fig:mixture}(b). 
\paragraph{LSIC-V3}
As shown in Figure \ref{fig:mixture}(c), we extend  LSIC-V2 by treating $\mathbf e^{u}_i$ (for user $i$) and $\mathbf e^{m}_j$ (for movie $j$)  as the static context vectors, and feed them as an extra input into the computation of the temporal hidden states of users and movies by LSTM. At each time step, the context information assists the inference of the hidden states of LSTM model. 
\paragraph{LSIC-V4}
This method use an attention mechanism to compute a weight for each hidden state by exploiting the global factors. The mixing scores at time $t$ can be reformulated by:
\begin{align}
r_{ij,t}&=g(\mathbf e^{u}_{i}, \mathbf e^{m}_{j}, \mathbf h^{u}_{i,t-1}, \mathbf h^{m}_{j,t-1}, \mathbf c^{u}_{i,t}, \mathbf c^{m}_{j,t})=\frac{1}{1+{\rm exp}(-\mathbf s)} \label{eq:soft}\\ 
\mathbf s&=\mathbf e^{u}_{i} \cdot \mathbf e^{m}_{j} + \mathbf h^{u}_{i,t} \cdot  \mathbf h^{m}_{j,t} +b_{i}+b_{j}\label{eq:soft-score}
\end{align}
where $\mathbf c^{u}_{i,t}$ and $\mathbf c^{m}_{j,t}$ are the context vectors at time step $t$ for user $i$ and movie $j$; $\mathbf h^{u}_{i,t}$ and $\mathbf h^{m}_{j,t}$ are the hidden states of LSTMs at time step $t$, computed by 

\begin{align}
\mathbf h^{u}_{i,t} &= {\rm LSTM}(\mathbf h^{u}_{i,t-1}, \mathbf z^{u}_{i,t}, \mathbf c^{u}_{i,t})\label{eq:lstm-soft-user}\\
\mathbf h^{m}_{j,t} &= {\rm LSTM}(\mathbf h^{m}_{j,t-1}, \mathbf z^{m}_{j,t}, \mathbf c^{m}_{j,t})\label{eq:lstm-soft-movie}
\end{align}

The context vectors $\mathbf c^{u}_{i,t}$ and $\mathbf c^{m}_{j,t}$ act as extra input in the computation of the hidden states in LSTMs to make sure that every time step of the LSTMs can get full information of the context (long-term information). 
The context vectors $\mathbf c^{u}_{i,t}$ and $\mathbf c^{m}_{j,t}$ are the dynamic representations of the relevant long-term information for user  $i$ and movie $j$ at time $t$, calculated by 
\begin{equation}
\label{eq:context}
\begin{aligned}
\mathbf c^{u}_{i,t}=\sum_{k=1}^U\alpha^{i}_{k,t} \mathbf e^{u}_{k} ;  \hspace{0.3cm} \mathbf c^{m}_{j,t}=\sum_{p=1}^M\beta^{j}_{p,t} \mathbf e^{m}_{p}
\end{aligned}
\end{equation}
where $U$ and $M$ are the number of users and movies. The attention weights $\alpha^{i}_{k,t}$ and $\beta^{j}_{p,t}$ for user $i$ and movie $j$  at time step $t$ are computed by 

\begin{align}
\alpha^{i}_{k,t}=\frac{\exp(\sigma(\mathbf h^{u}_{i,t-1}, \mathbf e^{u}_{k}))}{\sum_{k'=1}^U \exp(\sigma(\mathbf h^{u}_{i,t-1}, \mathbf e^{u}_{k'}))}\label{eq:lstm-attention-user}\\
\beta^{j}_{p,t}=\frac{\exp(\sigma(\mathbf h^{m}_{j,t-1}, \mathbf e^{m}_{p}))}{\sum_{p'=1}^M \exp(\sigma(\mathbf h^{m}_{j,t-1}, \mathbf e^{m}_{p'}))}\label{eq:lstm-attention-movie}
\end{align}
where  $\sigma$ is a feed-forward neural network to produce a real-valued score. The attention weights $\alpha^{i}_{t}$ and $\beta^{j}_{t}$  together determine which user and movie factors should be selected to generate $r_{ij,t}$. 

\subsection{Generative Adversarial Network (GAN) for Recommendation} 
Generative adversarial network (GAN)\cite{goodfellow2014generative} consist of a generator G and a discriminator D that compete in a  minimax game with two players: The discriminator tries to distinguish real high-rated movies on training data from ranking or recommendation list predicted by G, and the generator tries to fool the discriminator to generate(predict) well-ranked recommendation list. Concretely, D and G play the following game on V(D,G):
\begin{equation}
\label{eq:gan}
\begin{aligned}
\min_{G}\max_{D}v(D,G)=&\mathbb{E}_{x \sim P_{true}(x)}[{\rm log} D(x)] + \\
&\mathbb{E}_{z\sim P_{z}(z)}[{\rm log}(1-D(G(z)))]
\end{aligned}
\end{equation}
Here, $x$ is the input data from training set, $z$ is the noise variable sampled from normal distribution. 

We propose an  adversarial framework to iteratively optimize two models: the generative model $G$ predicting recommendation list given historical user-movie interactions, and the discriminative model $D$ predicting the relevance of the generated list. Like the standard generative adversarial networks (GANs)~\cite{goodfellow2014generative}, our model also optimizes the two models with a minimax two-player game.  $D$ tries to distinguish the real high-rated movies in the training data from the recommendation list generated by $G$, while $G$ maximizes the probability of $D$ making a mistake. Hopefully, this adversarial process can eventually adjust $G$ to generate plausible and high-quality recommendation list. We further elaborate the generator and discriminator below.


\vspace{-5pt}
\subsubsection{Discriminate Model} 
As depicted in Figure 1 (right side), we implement the discriminator $D$ via a Siamese Network that incorporates long and session-based ranking models in a pair-wise scenario. The discriminator $D$ has two symmetrical point-wise networks that share parameters and are updated by minimizing a pair-wise loss. 

The objective of discriminator $D$ is to maximize the probability of correctly distinguishing the ground truth movies from generated recommendation movies. For $G$ fixed, we can obtain the optimal parameters for the discriminator $D$ with the following formulation.
\vspace{-5pt}
\begin{equation}
\begin{aligned}
\label{eq:update_d}
\theta^*= \argmax_{\theta}\sum_{i \in  \mathcal{U}}\Big(\mathbb{E}_{m_{+},m_{-} \sim p_{true}}\left[{\log} D_{\theta}(u_{i},m_{-},m_{+}|t)\right]+ \\
\mathbb{E}_{m_{+}\sim p_{true},m_{g,t}\sim G_{\phi}(m_{g,t}|u_i,t)}\left[{\log}(1-D_{\theta}(u_{i},m_{g,t},m_{+}|t)\right]\Big)\\
\end{aligned}
\end{equation}
where $\mathcal{U}$ denotes the user set, $u_{i}$ denotes user $i$, $m_{+}$ is a positive (high-rating) movie, $m_{-}$ is a negative movie randomly chosen from the entire negative (low-rating) movie space, $\theta$ and $\phi$  are parameters of $D$ and $G$, and $m_{g,t}$ is the generated movie by  $G$ given time $t$. Here, we adopt hinge loss as our training objective since  it performs better than other training objectives. Hinge loss is widely adopted in various learning to rank scenario, which aims to penalize the examples that violate the margin constraint: 
\begin{equation}
\begin{aligned}
\label{eq:reward}
D(u_{i},m_{-}, m_{+}|t)=\max\Big\{0,\epsilon - 
g(\mathbf e^{u}_{i}, \mathbf e^{m}_{m_+}, \mathbf h^{u}_{i,t}, \mathbf h^{m}_{m_{+},t})\\
+g(\mathbf e^{u}_{i}, \mathbf e^{m}_{m_{-}}, \mathbf h^{u}_{i,t}, \mathbf h^{m}_{m_{-},t})\Big\}
\end{aligned}
\end{equation}
where $\epsilon$ is the hyper-parameter determining the margin of hinge loss, and we compress the outputs to the range of $(0,1)$.
\subsubsection{Generative Model} 

Similar to conditional GANs proposed in~\cite{mirza2014conditional}, our generator $G$ takes in the auxiliary information (user $u_{i}$ and time $t$) as input, and generates the ranking list for user $i$. Specifically, when D is optimized and fixed after computing Eq. \ref{eq:update_d}, the generator $G$ can be optimized by minimizing the following formulation:
\begin{equation}
\label{eq:update_g}
\begin{aligned}
\phi^* = \argmin_{\phi}\sum_{m \in \mathcal{M}}\big(\mathbb{E}_{m_{g,t} \sim G_{\phi}(m_{g,t}|u_i,t)}[{\rm log}(1-D(u_{i},m_{g,t},m_{+}|t)]\big)
\end{aligned}
\end{equation}
Here, $\mathcal{M}$ denotes the movie set. As in~\cite{goodfellow2014generative}, instead of minimizing ${\rm log}(1-D(u_{i},m_{g,t},m_{+}|t))$, we train $G$ to maximize ${\rm log}(D(u_{i},m_{g,t},m_{+}|t))$.  
\subsubsection{Policy Gradient}
Since the sampling of recommendation list by generator $G$ is discrete, it cannot be directly optimized by gradient descent as in the standard GAN formulation. Therefore, we use policy gradient based reinforcement learning algorithm~\cite{sutton2000policy} to optimize the generator $G$  so as to generate highly rewarded recommendation list. Concretely, we have the following derivations:
\begin{equation}
\begin{aligned} 
\label{eq:rl}
\nabla_{\phi} J^{G}(u_{i})&=\nabla_{\phi}\mathbb{E}_{m_{g,t}\sim G_{\phi}(m_{g,t}|u_i,t)}[\log D(u_{i},m_{g,t},m_{+}|t)] \\
&=\sum_{m \in \mathcal{M}}\nabla_{\phi}G_{\phi}(m|u_{i},t)\log D(u_{i},m,m_{+}|t)\\
&=\sum_{m \in \mathcal{M}} G_{\phi}(m|u_{i},t)\nabla_{\phi}\log G_{\phi}(m|u_{i},t)\log D(u_{i},m,m_{+}|t)\\
&=\mathbb{E}_{m_{g,t}\sim G_{\phi}(m|u_{i},t)}[\nabla_{\phi}{\rm log}G_{\phi}(m_{g,t}|u_{i},t)\log D(u_{i},m_{g,t},m_{+}|t)]\\
&\approx \frac{1}{K}\sum_{k=1}^K\nabla_{\phi}{\rm log}G_{\phi}(m_k|u_{i},t)\log D(u_{i},m_k,m_{+}|t)\\
\end{aligned}
\end{equation}
where $K$ is number of movies sampled by the current version of generator and  $m_k$ is the $k$-th sampled  item. With reinforcement learning terminology, we treat the term  ${\rm log}D(u_{i},m_{k},m_{+}|t)$ as the reward at time step $t$, and take an action $m_k$ at each time step. To accelerate the convergence, the rewards within a batch are normalized with a  Gaussian distribution to make the significant differences. 

\begin{algorithm}
  \label{alg:alg}
  \caption{Long and Session-based Ranking Model with Adversarial Network}
  \textbf{Input:} generator $G_{\phi}$, discriminator $D_{\theta}$, training data $S$.\\
  Initialize models $G_{\phi}$ and $D_{\theta}$ with random weights, and pre-train them on training data $S$.\\
  \textbf{repeat}\\
  	\For{g-steps}
	{
	    Generate recommendation list for user $i$ at time $t$ using the generator $G_{\phi}$.\\
	    Sample $K$ candidates from recommendation list.\\
	    \For{$k\in\{1,...,K\}$}
		{
        Sample a positive movie $m_{+}$ from S.\\
		Compute the reward ${\rm log}D(u_{i},m_{k},m_{+}|t)$ with Eq.(\ref{eq:reward})\\
		}
	    Update generator $G_{\phi}$ via policy gradient Eq.(\ref{eq:rl}).\\
	}
	\For{d-steps}
	{
		Use current $G_{\phi}$ to generate a negative movie and combined with a positive movie sampled from $S$.\\
	    Update discriminator $D_{\theta}$  with Eq.(\ref{eq:update_d}).\\
	}
  \textbf{until} convergence
\end{algorithm}

The overall procedure is summarized in Algorithm 1. During the training stage, the discriminator and the generator are trained alternatively in a adversarial manner via Eq.(\ref{eq:update_d}) and Eq.(\ref{eq:rl}), respectively.
\section{Experimental Setup}
\subsection{Datasets}

\begin{table}[h]
\small
\centering
\caption{Characteristics of the datasets.}
\label{tab:data}
	\resizebox{1.0\columnwidth}{!}
	{\footnotesize
	\begin{tabular}{l|lll}
		\toprule
		Dataset & Movielens-100K & Netflix-3M & Netflix-Full \\
		\midrule
		Users & 943 & 326,668& 480,189 \\
		movies &  1,6831 &17,751& 17,770 \\
		Ratings & 100,000  & 16,080,980 & 100,480,507\\
        \midrule
		Train Data &  09/97-03/98 &9/05-11/05 &  12/99-11/05\\
		Test Data & 03/98-04/98&  12/05 &  12/05\\
		Train Ratings & 77,714  & 13,675,402&  98,074,901\\
		Test Ratings & 21,875  &  2,405,578&  2,405,578\\
        \midrule
		Density  & 0.493 & 0.406  & 0.093\\
		Sparsity & 0.063 & 0.003  & 0.012\\
		\bottomrule
	\end{tabular}
	}
\end{table}
In order to evaluate the effectiveness of our model, we conduct experiments on two widely-used real-life datasets: Movielens100K  and Netflix (called ``Netflix-Full''). To evaluate the robustness of
our model, we also conduct experiments on a 3-month Netflix (called ``Netflix-3M'') dataset, which is a small version of Netflix-Full and has different training and testing period.
For each dataset, we split the whole data into several training and testing intervals based on time, as is done in~\cite{wu2017recurrent}, to simulate the actual situation of predicting future behaviors of users given the data that occurred strictly before current time step. Then, each testing interval is randomly divided into a validation set and a testing set. We removed the users and movies that do not appear in training set from the validation and test sets.  The detailed statistics are presented in Table \ref{tab:data}\footnote{``Density'' shows the average number of 5-ratings for the user per day. ``Sparsity'' shows the filling-rate of user-movie rating matrix as used in \cite{wu2017recurrent}}.  Following \cite{wang2017irgan}, we treat ``5-star'' in Netflix, ``4-start'' and ``5-start'' for  Movielens100K as positive feedback and all others as unknown (negative) feedback.

\subsection{Implementation Details}

\paragraph{Matrix Factorization.} We use matrix factorization with 5 and 16 factor numbers for Movielens and Netflix respectively \cite{wang2017irgan}. The parameters are randomly initialized by a uniform distribution ranged in [-0.05, 0.05].  We take gradient-clipping to suppress the gradient to the range of [-0.2,0.2]. L2 regularization (with $\lambda = 0.05$)  is used to the weights and biases of user and movie factors. 

\paragraph{Recurrent Neural Network.}  We use a single-layer LSTM with 10 hidden neurons,  15-dimensional input embeddings, and 4-dimensional dynamic states where each state contains 7-days  users/movies behavioral trajectories. That is, we take one month as the length of a session. The parameters are initialized with the same way as in MF. L2 regularization  (with $\lambda = 0.05$) is used to the weights and biases of the LSTM layer to avoid over-fitting.

\paragraph{Generative Adversarial Nets.} We pre-train $G$ and $D$ on the training data with a pair-wise scenario, and  use  SGD algorithm with learning rate $1 \times 10^{-4}$ to optimize its parameters. The number of sampled movies is set to 64 (i.e., $K=64$). In addition, we use matrix factorization model to generate 100 candidate movies, and then re-rank these movies with LSTM.  In all experiments, we conduct mini-batch training with batch size 128.\\
\subsection{Evaluation Metrics}
To quantitatively evaluate our method, we adopt the rank-based evaluation metrics to measure the performance of top-$n$ recommendation \cite{liu2008eigenrank,cremonesi2010performance} , including Precision@N, Normalised Discounted Cumulative Gain (NDCG@N), Mean Average Precision (MAP) and Mean Reciprocal Ranking (MRR).

\subsection{Comparison to Baselines}
In the experiments, we evaluate and compare our models with several state-of-the-art methods.\\
\begin{table*}[t]
	\centering
	\caption{Moive recommendation results (MovieLens). }\label{tab:cf-perf}
	\label{tab:movielens}
	\vspace{-8pt} 
	\resizebox{1.85\columnwidth}{!}{
	
	\begin{tabular}{l|c|c|c|c|c|c|c|c}
		\toprule
		 & \textbf{Precision@3} & \textbf{Precision@5} & \textbf{Precision@10}   & \textbf{NDCG@3} & \textbf{NDCG@5} & \textbf{NDCG@10} & \textbf{MRR} & \textbf{MAP} \\
		\midrule
        BPR & 0.2795 &  0.2664 &  0.2301  &  0.2910 &0.2761 & 0.2550 &0.4324 & 0.3549\\
        PRFM &  0.2884 & 0.2699 & 0.2481 & 0.2937 & 0.2894 & 0.2676 &0.4484 & 0.3885 \\
         LambdaFM &  0.3108 & 0.2953 & 0.2612 &  0.3302 & 0.3117 &0.2795 &0.4611 &  0.4014 \\ 
        RRN & 0.2893&   0.2740&   0.2480 & 0.2951 & 0.2814 & 0.2513 & 0.4320 & 0.3631 \\
		IRGAN &  0.3022 & 0.2885 & 0.2582 &  0.3285 & 0.3032 &0.2678 &0.4515 &  0.3744 \\  
       
        \midrule
        LSIC-V1 & 0.2946 &0.2713 & 0.2471&0.2905&  0.2801& 0.2644 & 0.4595 & 0.4066   \\
        LSIC-V2 & 0.3004 & 0.2843  & 0.2567 &0.3122 & 0.2951 & 0.2814 & 0.4624 & 0.4101\\
        LSIC-V3  &0.3105&   0.3023&   0.2610 &0.3217&   0.3086&  0.2912 & 0.4732& 0.4163\\
        LSIC-V4 & \textbf{0.3327} & \textbf{0.3173} & \textbf{0.2847} & \textbf{0.3512} & \textbf{0.3331} & \textbf{0.2939} & \textbf{0.4832} &  \textbf{0.4321} \\  		\midrule		
        Impv & 7.05\% & 7.45\% & 9.00\% & 6.36\% & 6.87\% & 5.15\% & 4.79\% & 7.65\%\\
		\bottomrule
	\end{tabular}
	}
\end{table*}

\begin{table*}[t]
	\centering
	\caption{Movie recommendation results (Netflix-3M). }\label{tab:netflix-3m}
	\resizebox{1.85\columnwidth}{!}{
	\vspace{-8pt} 
	\begin{tabular}{l|c|c|c|c|c|c|c|c}
		\toprule
		 & \textbf{Precision@3} & \textbf{Precision@5} & \textbf{Precision@10}   & \textbf{NDCG@3} & \textbf{NDCG@5} & \textbf{NDCG@10} & \textbf{MRR} & \textbf{MAP} \\
		\midrule
        BPR & 0.2670 & 0.2548 & 0.2403  &  0.2653 & 0.2576 & 0.2469 & 0.3829 & 0.3484 \\
        PRFM &  0.2562 & 0.2645 & 0.2661 &  0.2499 & 0.2575 &0.2614 &0.4022 &  0.3712 \\
        LambdaFM &  0.3082 & 0.2984 & 0.2812 &  0.3011 & 0.2993 &0.2849 &0.4316 &  0.4043 \\ 
        RRN & 0.2759 & 0.2741 & 0.2693 & 0.2685 & 0.2692 & 0.2676 & 0.3960& 0.3831  \\
		IRGAN &  0.2856 &0.2836 & 0.2715 &  0.2824 & 0.2813 & 0.2695 &  0.4060& 0.3718  \\  
        \midrule
        LSIC-V1 & 0.2815 & 0.2801 & 0.2680&0.2833 & 0.2742 & 0.2696  & 0.4416& 0.4025   \\
        LSIC-V2 & 0.2901 & 0.2883 & 0.2701 &0.2903 & 0.2831 & 0.2759 & 0.4406& 0.4102 \\
        LSIC-V3  &0.3152 & 0.3013 & 0.2722 &0.2927 & 0.2901 & 0.2821 & 0.4482& 0.4185\\
        LSIC-V4 & \textbf{0.3221} & \textbf{0.3193} & \textbf{0.2921} & \textbf{0.3157} & \textbf{0.3114} & \textbf{0.2975} &  \textbf{0.4501}& \textbf{0.4247}  \\  		\midrule		
        Impv & 4.51\% & 7.00\% & 3.88\% & 4.85\% & 4.04\% & 4.42\% & 4.29\% & 5.05\%\\
		\bottomrule
	\end{tabular}
	}
\end{table*}

\begin{table*}[t]
	\centering
	\caption{Movie recommendation results (Netflix-Full). }\label{tab:netflix-full}
		\vspace{-8pt} 
	\resizebox{1.85\columnwidth}{!}{
	
	\begin{tabular}{l|c|c|c|c|c|c|c|c}
		\toprule
		 & \textbf{Precision@3} & \textbf{Precision@5} & \textbf{Precision@10}  & \textbf{NDCG@3} & \textbf{NDCG@5} & \textbf{NDCG@10} & \textbf{MRR} & \textbf{MAP} \\
		\midrule
        BPR & 0.3011 & 0.2817 & 0.2587  &  0.2998 & 0.2870 & 0.2693 & 0.3840 & 0.3660 \\
        PRFM &  0.2959 & 0.2837 & 0.2624 &  0.2831 & 0.2887 & 0.2789 &0.4060 &  0.3916 \\
        LambdaFM &  0.3446 & 0.3301 & 0.3226 &  0.3450 & 0.3398 &0.3255 &0.4356 & 0.4067 \\ 
        RRN & 0.3135 & 0.2954 & 0.2699 & 0.3123 & 0.3004 & 0.2810 & 0.3953& 0.3768  \\
	IRGAN & 0.3320 &0.3229 & 0.3056 &  0.3319 &0.3260 & 0.3131 &  0.4248& 0.4052  \\  
        \midrule
        LSIC-V1 & 0.3127 & 0.3012 & 0.2818&0.3247 & 0.3098 & 0.2957  & 0.4470& 0.4098   \\
        LSIC-V2 & 0.3393 & 0.3271 & 0.3172 &0.3482 & 0.3401 & 0.3293 & 0.4448& 0.4213 \\
        LSIC-V3  &0.3501 & 0.3480 & 0.3291 &0.3498 & 0.3451& 0.3321 & 0.4503& 0.4257\\
        LSIC-V4 & \textbf{0.3621} & \textbf{0.3530} & \textbf{0.3341} & \textbf{0.3608} & \textbf{0.3511} & \textbf{0.3412}&  \textbf{0.4587}& \textbf{0.4327}  \\  		
        \midrule		
        Impv & 5.08\% & 6.94\% & 3.56\% & 4.58\% & 3.33\% & 4.82\% & 5.30\% & 6.39\%\\
		\bottomrule
	\end{tabular}
	}
\end{table*}
\begin{table*}[t]
	\centering
	\caption{Ablation results for Netflix-3M dataset. }\label{tab:ablation}
	\vspace{-8pt} 
	\resizebox{1.85\columnwidth}{!}{
	
	\begin{tabular}{l|c|c|c|c|c|c|c|c}
		\toprule
		 &\textbf{Precision@3} & \textbf{Precision@5} & \textbf{Precision@10}  & \textbf{NDCG@3} & \textbf{NDCG@5} & \textbf{NDCG@10} & \textbf{MRR} & \textbf{MAP} \\
		\midrule
       LSIC-V4 & \textbf{0.3221} & \textbf{0.3193} & \textbf{0.2921} & \textbf{0.3157} & \textbf{0.3114} & \textbf{0.2975} &  \textbf{0.4501}& \textbf{0.4247}  \\ 
       w/o RL & 0.3012 & 0.2970 & 0.2782 &0.2988 & 0.2927 & 0.2728 & 0.4431& 0.4112 \\
       w/o poster  &0.3110 & 0.3012 & 0.2894 & 0.3015 & 0.3085 & 0.2817 & 0.4373& 0.4005\\

		\bottomrule
	\end{tabular}
	}
\end{table*}

\begin{table*}[t]
 \label{tab:cases}
 	\small
 	\centering
 	\caption{ The Recalled movies from Top-10 candidates in Neflix-3M dataset }\label{tab:cf-perf}
	\vspace{-8pt} 
 	\resizebox{2.2\columnwidth}{!}{	
 	\begin{tabular}{l|l|l|l|l|l|l|l}
 		\toprule
 		  &\textbf{Groundtruth} & \textbf{IRGAN}~\cite{wang2017irgan} & \textbf{RRN}~\cite{wu2017recurrent} & \textbf{LambdaFM}~\cite{yuan2016lambdafm} & \textbf{LSIC-V4}  \\
 		\midrule
 	Userid: 1382
		& \makecell [l]{  9 Souls\\
					The Princess Bride\\
					Stuart Saves His Family\\
					The Last Valley\\
					Wax Mask\\
					After Hours \\
					Session 9\\
					Valentin \\
				} 
         & \makecell[l]{  \textbf{[1]} The Beatles: Love Me Do \\
					\textbf{[2]} Wax Mask \checkmark \\ 
					\textbf{[3]} Stuart Saves His Family \checkmark \\ 
					\midrule 
					\textbf{[4]} After Hours \checkmark \\
					\textbf{[5]} Top Secret! \\
					\midrule 
					\textbf{[6]} Damn Yankees \\
					\textbf{[7]} Dragon Tales: It's Cool to Be Me! \\
					\textbf{[8]} Play Misty for Me \\
					\textbf{[9]} The Last Round: Chuvalo vs. Ali' \\
					\textbf{[10]} La Vie de Chateau \\
				}
		&  \makecell[l]{ \textbf{[1]} Falling Down \\
					\textbf{[2]} 9 Souls \checkmark\\ 
					\textbf{[3]} Wax Mask \checkmark\\ 
					\midrule 
					\textbf{[4]} After Hours \checkmark\\ 
					\textbf{[5]} Stuart Saves His Family \checkmark \\
					\midrule 
					\textbf{[6]} Crocodile Dundee 2 \\
					\textbf{[7]} The Princess Bride  \checkmark \\
					\textbf{[8]} Dragon Tales: It's Cool to Be Me! \\
					\textbf{[9]} They Were Expendable \\
					\textbf{[10]} Damn Yankees\\					
				}
		  & \makecell[l]{\textbf{[1]} The Avengers '63 \\ 
		  			\textbf{[2]} Wax Mask \checkmark \\ 
					\textbf{[3]} The Boondock Saints \\ 
					\midrule 
					\textbf{[4]} Valentin \checkmark\\
					\textbf{[5]} 9 Souls \checkmark \\
					\midrule 
					\textbf{[6]} The Princess Bride \checkmark \\
					\textbf{[7]} After Hours \checkmark \\
					\textbf{[8]} Tekken \\
					\textbf{[9]} Stuart Saves His Family \checkmark \\
					\textbf{[10]} Runn Ronnie Run 
				}		
		  
		  & \makecell[l]{\textbf{[1]}9 Souls \checkmark \\ 
		  			\textbf{[2]}The Princess Bride \checkmark\\ 
					\textbf{[3]}Stuart Saves His Family \checkmark\\ 					
					\midrule 
					\textbf{[4]} The Last Valley \checkmark \\ 
					\textbf{[5]} Wax Mask \checkmark \\ 
					\midrule 
					\textbf{[6]} Session 9 \checkmark \\ 
					\textbf{[7]} Dragon Tales: It's Cool to Be Me! \\ 
					\textbf{[8]} Damn Yankees\\
					\textbf{[9]} After Hours \checkmark \\ 
					\textbf{[10]} Valentin \checkmark
				}\\	
	\midrule	  
         Userid: 8003
         & \makecell[l]{9 Souls \\ Princess Bride} 
         & \makecell [l]{ \textbf{[1]} Cheech Chong's Up in Smoke\\
         			\textbf{[2]} Wax Mask \\ 
				\textbf{[3]} Damn Yankees\\
				\midrule 
				\textbf{[4]} Dragon Tales: It's Cool to Be Me! \\
				\textbf{[5]} Top Secret!\\
                \midrule
				\textbf{[6]} Agent Cody Banks 2: Destination London\\
				\textbf{[7]} After Hours\\
				\textbf{[8]} Stuart Saves His Family \\
				\textbf{[9]} 9 Souls \checkmark\\
				\textbf{[10]} The Beatles: Love Me Do \\
			}      
         & \makecell [l]{ \textbf{[1]} Crocodile Dundee 2\\
			         \textbf{[2]} Session 9\\
				\textbf{[3]} Falling Down\\
				\midrule 
				\textbf{[4]} Wax Mask \\
				\textbf{[5]} After Hours\\
                \midrule
				\textbf{[6]} Stuart Saves His Family\\
				\textbf{[7]} 9 Souls \checkmark\\ 
				\textbf{[8]} The Princess Bride  \checkmark\\
				\textbf{[9]} Dragon Tales: It's Cool to Be Me!\\
				\textbf{[10]} Scream 2\\
			} 

         & \makecell [l]{ \textbf{[1]} The Insider\\
         			\textbf{[2]}A Nightmare on Elm Street 3\\ 
				\textbf{[3]} Dennis the Menace Strikes Again\\
				\midrule 
				\textbf{[4]} Civil Brand \\
				\textbf{[5]} 9 Souls \checkmark\\
                \midrule
				\textbf{[6]} Falling Down\\
				\textbf{[7]} The Princess Bride \checkmark\\
				\textbf{[8]} Radiohead: Meeting People\\
				\textbf{[9]} Crocodile Dundee 2\\
				\textbf{[10]} Christmas in Connecticut\\
			} 
         & \makecell[l]{\textbf{[1]} 9 Souls \checkmark \\ 
         			\textbf{[2]}The Princess Bride \checkmark \\ 
				\textbf{[3]}The Last Valley \\ 
				\midrule
				\textbf{[4]}Stuart Saves His Family \\ 
				\textbf{[5]}Wax Mask \\ 
				\midrule
				\textbf{[6]}Dragon Tales: It's Cool to Be Me! \\ 
				\textbf{[7]}Session 9 \\ 
				\textbf{[8]}Crocodile Dundee 2 \\
				\textbf{[9]}Damn Yankees \\
				\textbf{[10]}Cheech Chong's Up in Smoke
			} \\
 
          \bottomrule
 	\end{tabular}
 	}   
 \end{table*}
\vspace{-5pt}
\paragraph{Bayesian Personalised Ranking (BPR)} Given a positive movie, BPR uniformly samples negative movies to resolve the imbalance issue and provides a basic baseline for top-N recommendation ~\cite{rendle2009bpr}.

\paragraph{Pairwise Ranking Factorization Machine (PRFM)}  This is one of the state-of-the-art movie recommendation algorithms, which applies Factorization Machine model to
microblog ranking on basis of pairwise classification \cite{qiang2013exploiting}.  We use the same settings as in \cite{qiang2013exploiting} in our experiments. 
\paragraph{LambdaFM} It is a strong baseline for recommendation, which directly optimizes the rank biased metrics~\cite{yuan2016lambdafm}.  We run the LambdaFM model with the publicly available code\footnote{https://github.com/fajieyuan/LambdaFM}, and use default settings for all hyperparameters. 

\paragraph{Recurrent Recommender Networks (RRN)}  This model supplements matrix factorization with recurrent neural network model via a hard mixture mechanism~\cite{wu2017recurrent}. We use the same setting as in~\cite{wu2017recurrent}. 

\paragraph{IRGAN} This model trains the generator and discriminator alternatively with MF in an adversarial process ~\cite{wang2017irgan}. 
We run the IRGAN model with the publicly available code\footnote{https://github.com/geek-ai/irgan}, and use default settings for all hyperparameters. 

\section{Experimental Results}
In this section, we compare our model with baseline methods quantitatively and qualitatively. 

\subsection{Quantitative Evaluation}

We first evaluate the performance of top-$n$ recommendation. The experimental results are summarized in Tables \ref{tab:movielens} ,\ref{tab:netflix-3m} and \ref{tab:netflix-full}. Our model substantially and consistently outperforms the baseline methods by a noticeable margin on all the experimental datasets. In particular, we have explored several versions of our model with different mixture mechanisms. As one anticipates,  LSIC-V4 achieves the best results across all evaluation metrics and all datasets. For example, on MovieLens dataset, LSIC-V4  improves  $7.45\%$ on percision@5 and $6.87\%$ on NDCG@5  over the baseline methods. The main strength of our model comes from its capability of prioritizing both long-term and short-term information in content-aware movie recommendation. In addition, our mixture mechanisms (hard and soft) also seem quite effective to integrate MF and RNN.

To better understand the adversarial training process,  we visualize the learning curves of LSIC-V4 as shown in Figure 3. Due to the limited space, we only report the Pecision@5 and NDCG@5 scores  as in~\cite{wang2017irgan}. The other metrics exhibit a similar trend. As shown in Figure 3, after about 50 epoches, both Precision@5 and NDCG@5 converge and the winner player is the generator which is used to generate recommendation list for our final top-$n$ movie recommendation. 
The performance of generator $G$ becomes better with the effective feedback (reward) from discriminator $D$. 
On the other hand, once we have a set of high-quality recommendation movies, the performance of $D$ deteriorates gradually in the training procedure and makes mistakes for predictions.
In our experiments, we use the generator G with best performance to predict test data.

\subsection{Ablation Study}

In order to analyze the effectiveness of different components of our model for  top-$n$ movie recommendation, in this section, we report the ablation test of LSIC-V4 by discarding poster information (w/o poster)  and replacing the reinforcement learning with Gumbel-Softmax \cite{kusner2016gans} (w/o RL), respectively.  Gumbel-Softmax is an alternative method to address the non-differentiation problem so that $G$ can be trained straightforwardly.

Due to the limited space, we only illustrate the experimental results for Netflix-3M dataset that is widely used in movie recommendation (see Table \ref{tab:ablation}).  Generally, both factors contribute, and reinforcement learning contributes most. This is within our expectation since discarding reinforcement learning will lead the adversarial learning inefficient. With Gumbel-Softmax, $G$ does not benefit from the reward of $D$, so that we do not know which movies sampled by G are good and should be reproduced. Not surprisingly, poster information also contributes to movie recommendation.

\subsection{Case Study}
In this section, we will further show the advantages of our models through some quintessential examples.

In Table 7, we provide the recommendation lists generated by three state-of-the-art baseline methods (i.e., IRGAN, RNN, LambdaFM) as well as the proposed  LSIC-V4 model for two users who are randomly selected from the Netflix-3M dataset. Our model can rank the positive movies in higher positions than other methods. For example, the ranking of the movie ``\emph{9 souls}'' for user ``8003'' has increased from 5-th position (by LambdaFM) to 1st position (by LSIC-V4). Meanwhile some emerging movies such as ``\emph{Session 9}'' and ``\emph{The Last Valley}'' that are truly attractive to the user ``1382'' have been recommended by our models, whereas they are ignored by baseline methods. In fact, we can include all positive movies in the top-10 list for user ``1382'' and in  top-3 list for user ``8003''. Our model benefits from the fact that both dynamic and static knowledge are incorporated into the model with adversarial training.

\begin{figure}[h]
\centering
\subfigure{\includegraphics[width=0.49\columnwidth]{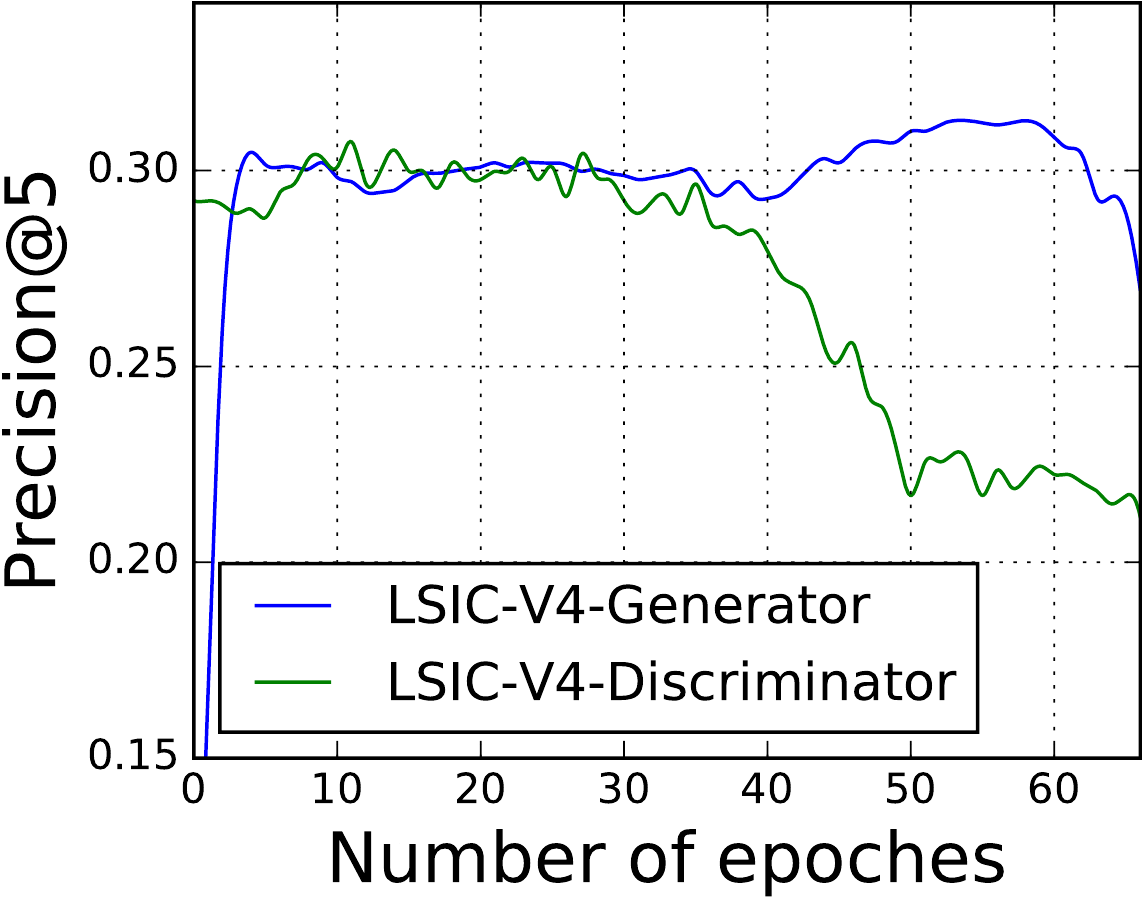}\label{fig:percision5}}
\subfigure{\includegraphics[width=0.49\columnwidth]{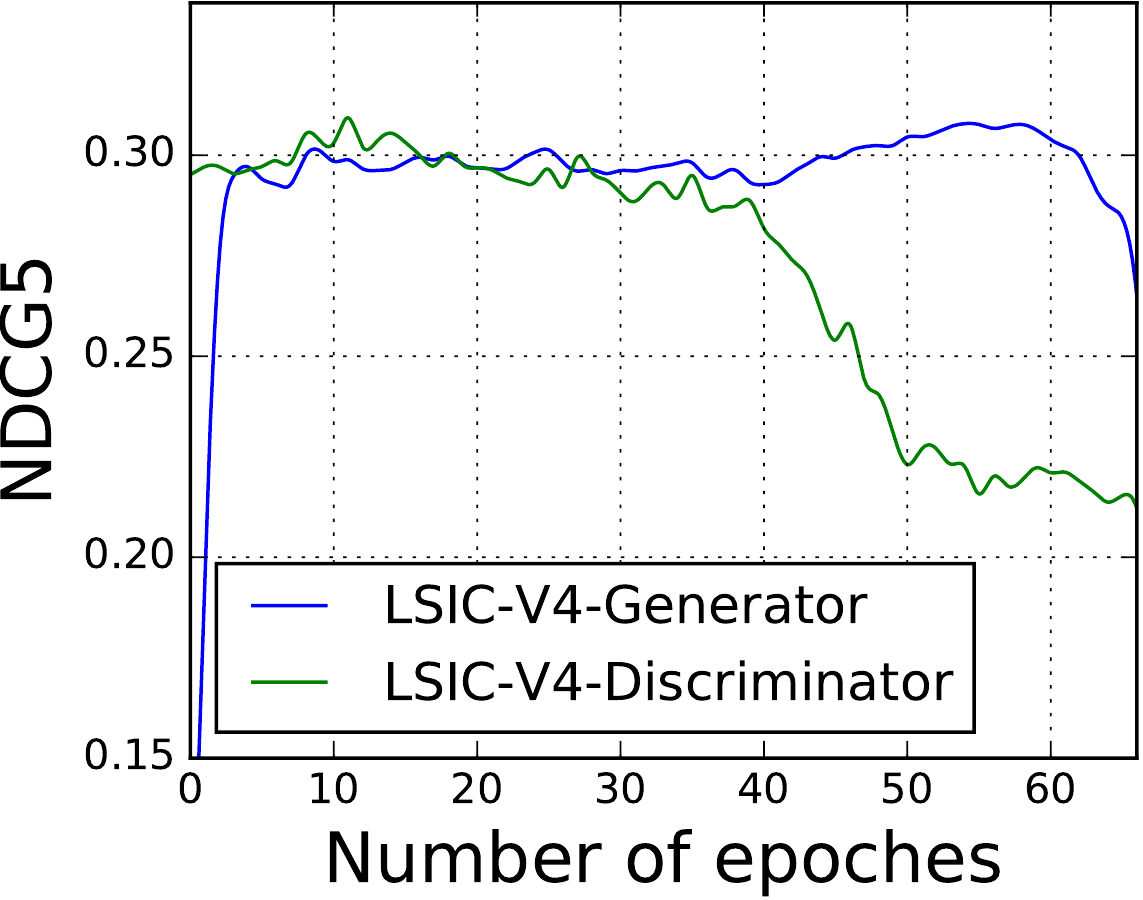}\label{fig:curveMRRquadratic}}
\caption{Learning curves of GAN on Netflix-3M}\label{fig:mrr}
\vspace{-5pt}
\end{figure}

\subsection{Re-rank Effect}
\begin{figure}[h]
\centering
\subfigure{\includegraphics[width=0.49\columnwidth]{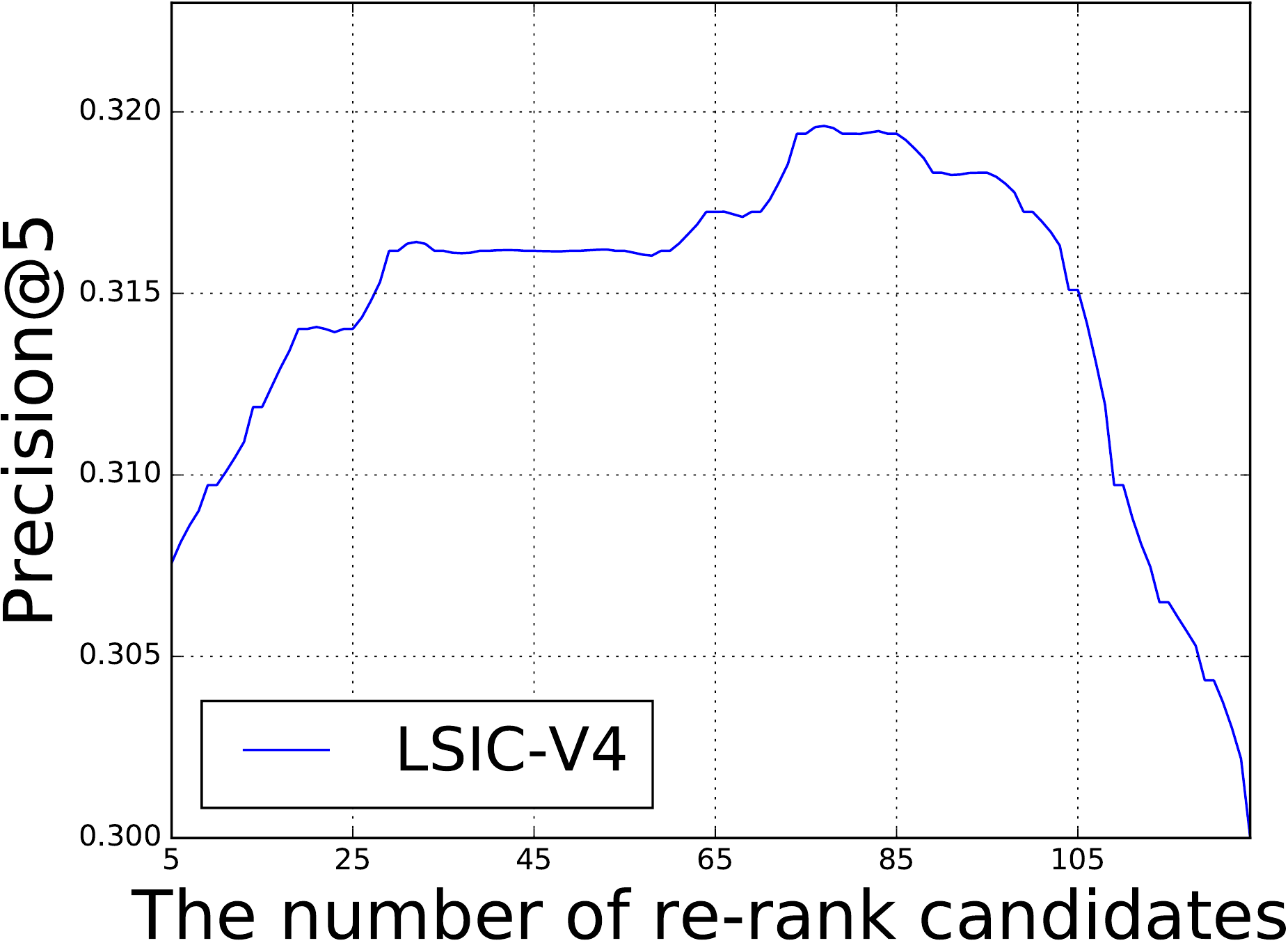}\label{fig:p5_rerank}}
\subfigure{\includegraphics[width=0.49\columnwidth]{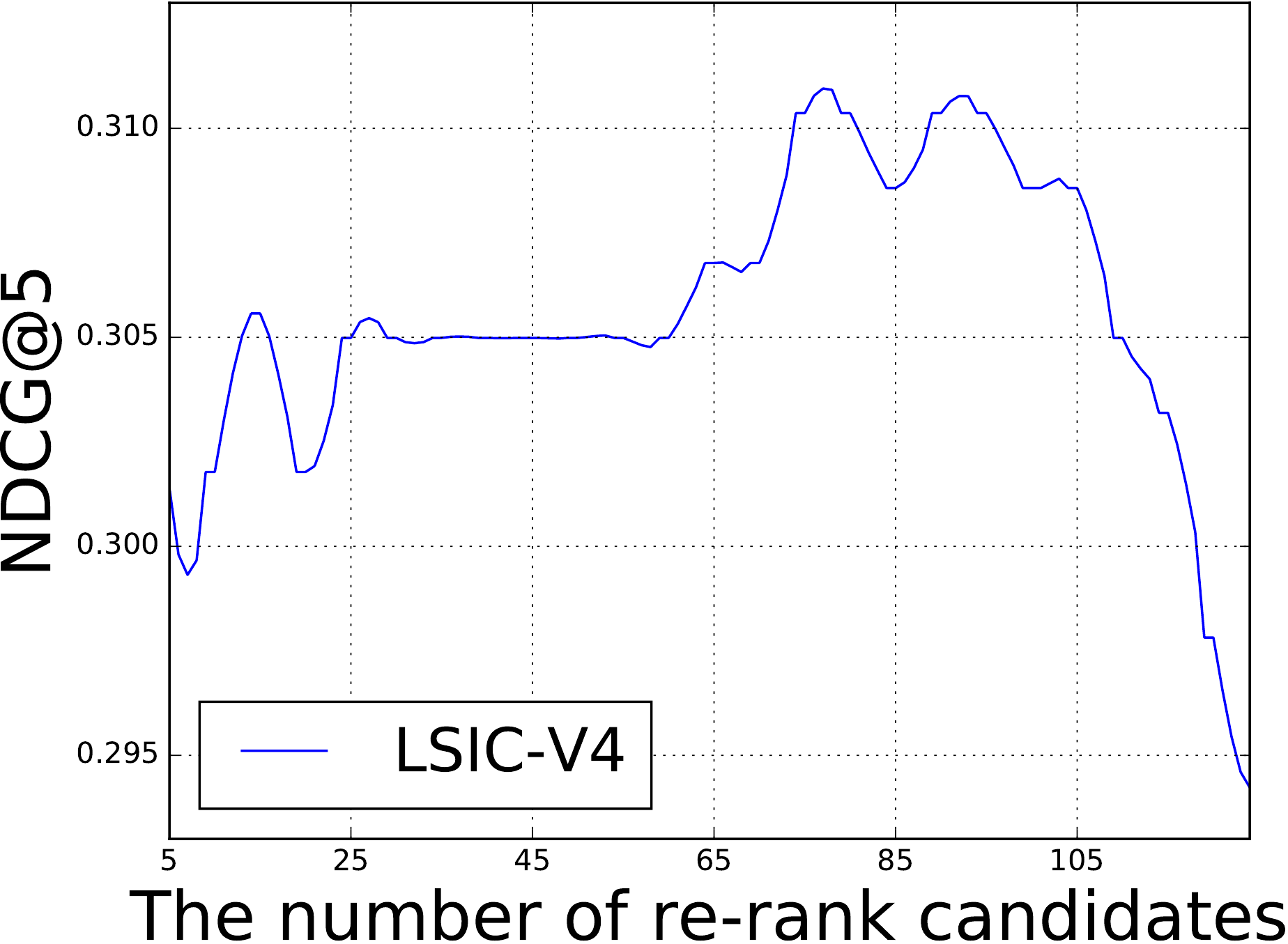}\label{fig:ndcg5_rerank}}
\caption{Sensitivity of the candidate scale on Netflix-3M}\label{fig:rerank}
\vspace{-5pt}
\end{figure}

From our experiments, we observe that it could be time-consuming for RNN to infer all movies. In addition, users may be interested in a few movies that are subject to a long-tailed distribution. Motivated by these observations, we provide a re-ranking strategy as used in \cite{covington2016deep}. Specifically, we first generate $N$ candidate movies by MF, and then re-rank these candidate movies with our model. In this way, the inference time can been greatly reduced. Figure \ref{fig:rerank} illustrates the performance curves over the number of candidate movies (i.e. $N$) generated by MF. We only report the Pecision@5 and NDCG@5 results due to the limited space, the other metrics  exhibit a similar trend.  As shown in Figure \ref{fig:rerank}, when the number of candidate movies is small, i.e., $N\leq 100$ for Netflix-3M dataset, the Percision@5 raises gradually with the number of candidate movies increases. Nevertheless, the performance drops rapidly with the increasing candidate movies when $N\geq 100$. It suggests that the candidates in a long-tail side generated by MF is inaccurate and these candidate movies deteriorate the overall performance of the re-rank strategy. 


\subsection{Session Period Sensitivity}
\begin{figure}[h]
\centering
\subfigure{\includegraphics[width=0.49\columnwidth]{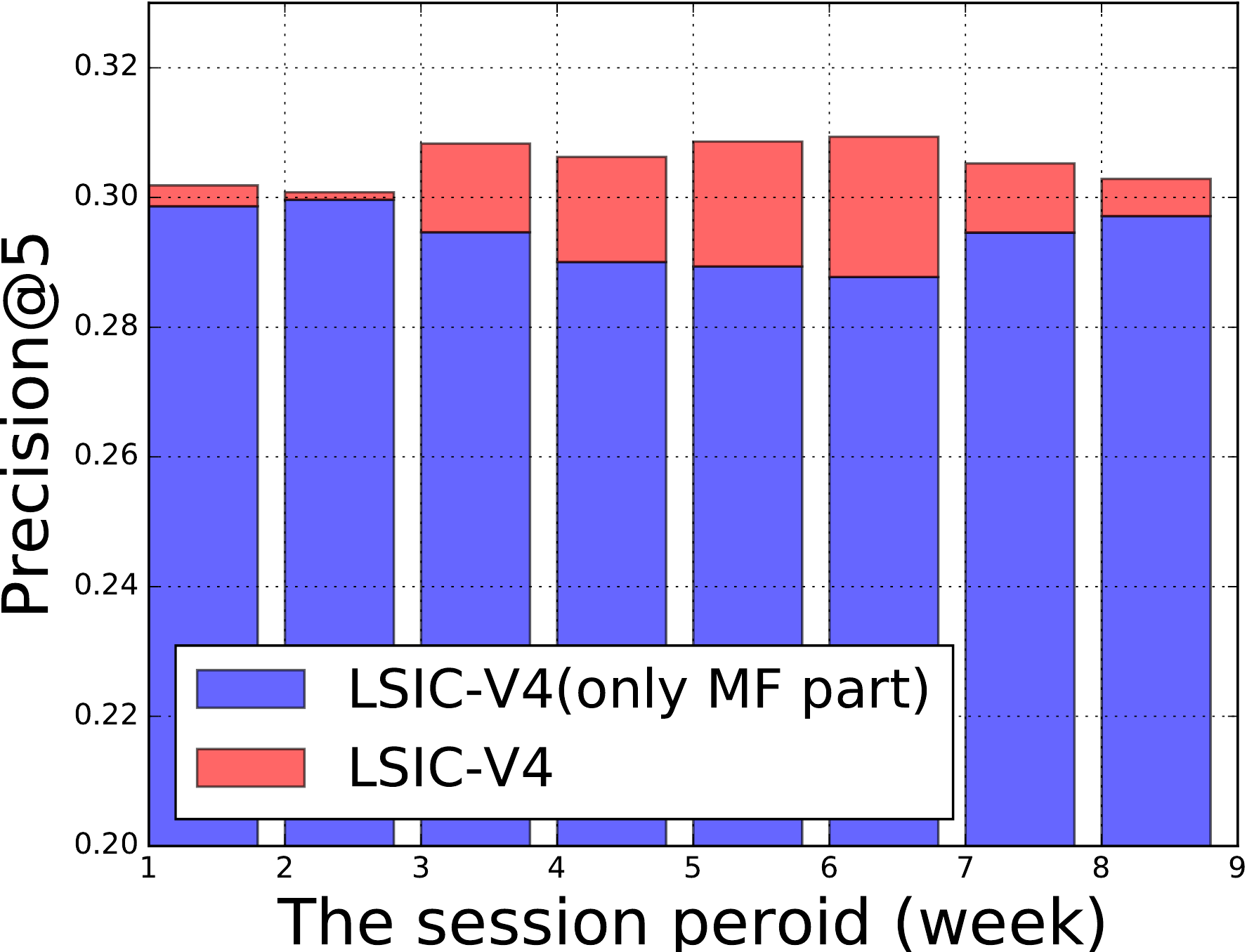}}
\subfigure{\includegraphics[width=0.49\columnwidth]{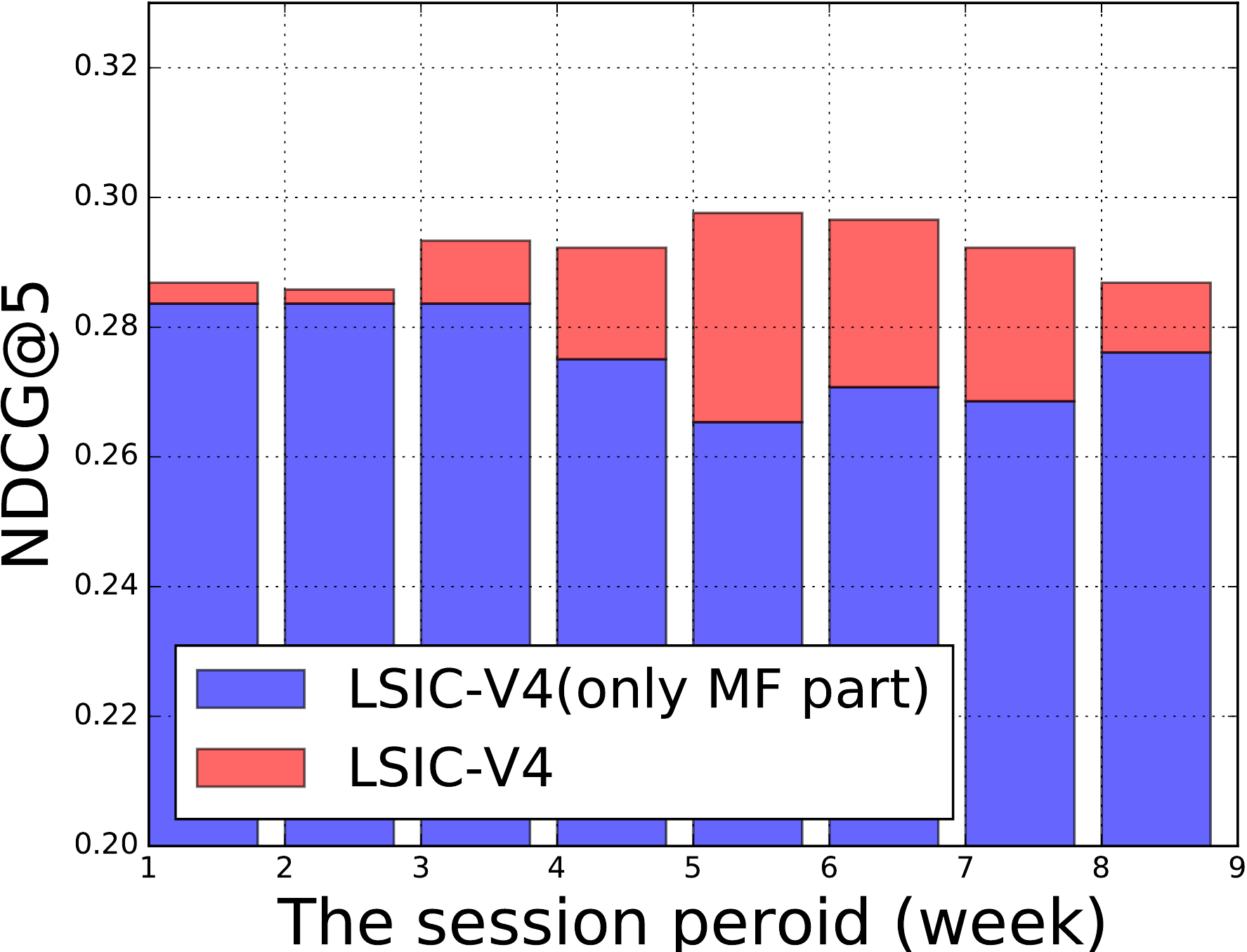}}
\caption{Sensitivity of the session period on Netflix-3M}\label{fig:session}
\vspace{-5pt}
\end{figure}

The above experimental results have shown that the session-based (short-term) information indeed improves the performance of top-N recommendation. We conduct an experiment on Netflix-3M to investigate how the session period influences the final recommendation performance. As shown in Figure \ref{fig:session}, the bars in purple color (below) describe the basic performance of MF component while the red ones (above) represent the extra improvement by the LSIC-V4. The RNN plays an insignificant role in the very early sessions since it lacks enough historical interaction data. For later period of sessions, RNN component tends to be more effective, and our joint model achieves a clear improvement over the model with only MF component.  

\vspace{-5pt}


 \section{Conclusion}
 In this paper, we proposed a novel adversarial process for top-$n$ recommendation.
Our model incorporated both matrix factorization and recurrent neural network to exploit the benefits of the long-term and short-term knowledge. We also integrated poster information to further improve the performance of movie recommendation.  Experiments on two real-life datasets showed the performance superiority of our model.

\bibliographystyle{ACM-Reference-Format}
\bibliography{2018-www-recommendation}
\end{document}